\documentclass[preprint,12pt]{elsarticle}

\usepackage{amssymb}
\usepackage{amsmath}
\usepackage{booktabs}
\usepackage{listings}
\usepackage{xcolor}
\usepackage[hyphens]{url}
\usepackage{hyperref}
\usepackage{cleveref}
\usepackage[T1]{fontenc}
\usepackage{tabularx}
\usepackage{textcomp}
\usepackage{microtype}

\newdefinition{definition}{Definition}

\definecolor{haskellblue}{rgb}{0.1, 0.1, 0.6}
\definecolor{commentgreen}{rgb}{0.15, 0.4, 0.15}
\definecolor{stringred}{rgb}{0.6, 0.1, 0.1}
\definecolor{codegray}{rgb}{0.4, 0.4, 0.4}
\definecolor{backcolor}{rgb}{0.97, 0.97, 0.97}

\lstdefinestyle{HaskellClean}{
    language=Haskell,
    basicstyle=\small\ttfamily,
    keywordstyle=\color{haskellblue}\bfseries,
    commentstyle=\color{commentgreen},
    stringstyle=\color{stringred},
    showstringspaces=false,
    breaklines=true,
    backgroundcolor=\color{backcolor},
    frame=single,
    framerule=0.4pt,
    rulecolor=\color{codegray},
    numbers=left,
    numberstyle=\tiny\color{codegray},
    upquote=true,
    tabsize=2,
    captionpos=b,
    xleftmargin=2em,
    framexleftmargin=1.5em,
    aboveskip=1em,
    belowskip=1em,
    morekeywords={class, instance, type, Config, StatisticalProtocol,
                  isValidTransition, DataContract, StatisticalTestSpec,
                  ExceptT, throwError, data, where, lift, StateT, liftIO,
                  MonadIO, Either, Right, Left, Maybe, Just, Nothing}
}

\lstdefinestyle{PythonClean}{
    language=Python,
    basicstyle=\small\ttfamily,
    keywordstyle=\color{haskellblue}\bfseries,
    commentstyle=\color{commentgreen},
    stringstyle=\color{stringred},
    showstringspaces=false,
    breaklines=true,
    backgroundcolor=\color{backcolor},
    frame=single,
    framerule=0.4pt,
    rulecolor=\color{codegray},
    numbers=left,
    numberstyle=\tiny\color{codegray},
    upquote=true,
    tabsize=4,
    captionpos=b,
    xleftmargin=2em,
    framexleftmargin=1.5em,
    aboveskip=1em,
    belowskip=1em
}

\definecolor{adablue}{rgb}{0.0, 0.2, 0.6}

\lstdefinestyle{AdaClean}{
    language=Ada,
    basicstyle=\small\ttfamily,
    keywordstyle=\color{adablue}\bfseries,
    commentstyle=\color{commentgreen},
    stringstyle=\color{stringred},
    showstringspaces=false,
    breaklines=true,
    backgroundcolor=\color{backcolor},
    frame=single,
    framerule=0.4pt,
    rulecolor=\color{codegray},
    numbers=left,
    numberstyle=\tiny\color{codegray},
    upquote=true,
    tabsize=3,
    captionpos=b,
    xleftmargin=2em,
    framexleftmargin=1.5em,
    aboveskip=1em,
    belowskip=1em,
    mathescape=false,
    literate={'}{\textquotesingle}1,
    morekeywords={pragma,Assert,SPARK_Mode,Long_Float,Loop_Invariant,Post}
}

\definecolor{leanpurple}{rgb}{0.35, 0.1, 0.5}

\lstdefinestyle{LeanClean}{
    basicstyle=\small\ttfamily,
    keywordstyle=\color{leanpurple}\bfseries,
    commentstyle=\color{commentgreen},
    showstringspaces=false,
    breaklines=true,
    backgroundcolor=\color{backcolor},
    frame=single,
    framerule=0.4pt,
    rulecolor=\color{codegray},
    numbers=left,
    numberstyle=\tiny\color{codegray},
    upquote=true,
    tabsize=2,
    captionpos=b,
    xleftmargin=2em,
    framexleftmargin=1.5em,
    aboveskip=1em,
    belowskip=1em,
    keywords={def,theorem,noncomputable,Prop,Measure,Set,Finset,fun,if,then,
              else,max,MeasurableSpace,MeasurableSet,Measurable,
              IsProbabilityMeasure,IsUniformPValue,IsIndepOfSubalgebra},
    literate=
      {μ}{{$\mu$}}1 {Ω}{{$\Omega$}}1 {ω}{{$\omega$}}1 {α}{{$\alpha$}}1
      {ε}{{$\varepsilon$}}1 {→}{{$\rightarrow$}}2 {≤}{{$\leq$}}2
      {∀}{{$\forall$}}1 {∃}{{$\exists$}}1 {∈}{{$\in$}}1 {∩}{{$\cap$}}1 {∑}{{$\textstyle\sum$}}1
      {∫}{{$\textstyle\int$}}1 {∂}{{$\partial$}}1 {ᵐ}{{$^{\mathrm{ae}}$}}1
      {ℝ}{{$\mathbb{R}$}}1 {ℕ}{{$\mathbb{N}$}}1 {₀}{{$_0$}}1
      {⁻¹}{{$^{-1}$}}2 {≠}{{$\neq$}}2 {⊆}{{$\subseteq$}}2
}

\begin{document}

\journal{Journal of Logical and Algebraic Methods in Programming}

\begin{frontmatter}

\title{Structural Enforcement of Statistical Rigor in AI-Driven Discovery: A Functional Architecture}

\author[sinica]{Karen Sargsyan\corref{cor1}}
\ead{karen.sarkisyan@gmail.com}
\cortext[cor1]{Corresponding author}

\affiliation[sinica]{organization={Institute of Chemistry, Academia Sinica},
            city={Taipei},
            country={Taiwan}}

\begin{abstract}
AI-Scientist systems risk manufacturing spurious discoveries through uncontrolled multiple testing. We present a functional architecture that enforces statistical rigor at two levels: a Haskell embedded domain-specific language (the Research monad) that makes it impossible to test a hypothesis without updating the error budget, and a declarative scaffold that fixes the data flow and the statistical test, together with an OS-level sandbox that makes validation data physically absent from the environment in which LLM-generated code runs. We treat FDR control as a formal requirement and trace it to the implementation. We ground the design in a machine-checked Lean~4 formalization of LORD online false-discovery-rate (FDR) control: we derive its error budget and prove marginal FDR control, and full FDR control when thresholds do not adapt to earlier rejections. We then verify in SPARK/Ada that the LORD thresholds, computed in IEEE~754 arithmetic, never exceed the available wealth, given a margin condition that our configurations meet by a factor of at least eight; without a margin the property fails. To our knowledge this is the first machine-checked proof of an online FDR control theorem. In simulation, the architecture holds the false discovery rate near 1\% against a 5\% target, where a naive approach reaches 41\%. In end-to-end case studies, a valid test avoids the false discoveries a flawed one produces, yet still finds real effects when the data allow. An adversarial evaluation confirms that, inside the sandbox, generated code cannot read the held-out data even when given its exact path.
\end{abstract}

\begin{keyword}
false discovery rate \sep formal verification \sep
functional programming \sep AI-Scientist \sep
SPARK/Ada \sep Lean 4 \sep requirements engineering
\end{keyword}

\end{frontmatter}

\section{Introduction}

Suppose an automated research system tests two thousand hypotheses
overnight. At significance level $\alpha = 0.05$, roughly a hundred
will appear significant by chance alone---even if none of the
hypotheses are true. This concern is not hypothetical. AI-Scientist
systems~\cite{sakana2024aiscientist,schmidgall2025agent,wei2025agentic,zhang2025deep,swanson2025virtual}
that use large language models (LLMs) to generate, implement, and
evaluate scientific hypotheses are already producing manuscripts that
reach the acceptance threshold of workshop peer
review~\cite{yamada2025ai}, and tools to scale them are under active
development~\cite{gao2025democratizing}.
As these systems mature, the integrity of their
output, particularly their statistical claims, becomes a pressing
concern~\cite{kon2025curie,zhu2025safescientist}.

Two distinct challenges arise. The first is the classical multiple
testing problem: testing many hypotheses at one fixed threshold lets
the false discovery rate (FDR)---the fraction of reported findings that
are false---climb far above the level one intended. Online FDR control
methods like
LORD (Levels based On Recent Discovery)~\cite{javanmard2018online}
solve this mathematically, but their guarantees rest on assumptions that an
implementation can silently violate. The second is
architectural: AI-Scientists operate in hybrid environments where a
functional orchestrator dispatches work to LLM-generated Python code.
At this trust boundary, an LLM might inadvertently use validation data
during model training, producing artificially low p-values that
defeat any statistical correction.

We address both challenges through a defense-in-depth architecture.
At the macro level, the \texttt{Research} monad---a Haskell embedded
DSL---makes it structurally impossible to test a hypothesis without
correctly updating the statistical state. At the micro level,
declarative scaffolding generates rigid Python harnesses that fix the
data flow and the statistical test, and an OS-level sandbox, which we
implement and test as a standalone component, runs the optimization
step in an environment in which the validation data is physically
absent, so the optimization code cannot read it by any path. To establish that these structural properties suffice, we
formalize LORD in the Lean~4 proof assistant: we derive its error
budget from the threshold formula and machine-check marginal FDR
control and, for thresholds that do not adapt to earlier rejections,
full FDR control, isolating four conditions an implementation must
meet. Three are structural---information flow, data separation, and
test validity---enforced by the types and the scaffold; the fourth is
arithmetic, that the wealth process stay non-negative under
floating-point rounding, which we verify over IEEE~754 doubles in
SPARK/Ada for the LORD thresholds themselves, given a margin
condition that we show cannot be dropped. To our
knowledge this is the first machine-checked proof of an online FDR
control theorem.

Read as a requirements problem, the paper follows one requirement
from its statement to its enforcement. The top-level requirement is
that the false discovery rate of reported findings stays at or below a
target level at every fixed horizon. The Lean theorem turns it into
four derived requirements, (H1)--(H4), each a hypothesis of a
machine-checked theorem, and we trace each one to the mechanism that
enforces it and to the evidence that it holds. Not every link in this
chain is a proof: some requirements are enforced by construction,
meaning a program that violates them does not typecheck or never
receives the data; one is proved over the real numbers and, under a
margin condition, over floating point; and one is an
assumption supported by experiment. A traceability matrix
(Table~\ref{tab:conditions}) records which is which.

We validate the approach through Monte Carlo simulation, end-to-end
case studies, and an adversarial evaluation of the data-separation
boundary. A running example---an AI-Scientist optimizing SVM
classifiers on the Wine dataset~\cite{wine_dataset}---runs throughout
the paper.

\section{Background: Online FDR Control}
\label{sec:background}

This section develops the statistical machinery our architecture
enforces. Readers familiar with online FDR control may skip to
Section~\ref{sec:example}.

\subsection{The Multiple Testing Problem}

Testing a single hypothesis at level $\alpha = 0.05$ accepts a 5\%
chance of a false positive. Testing $m$ hypotheses independently at
the same level raises the probability of at least one false
positive to $1 - (1-\alpha)^m$. For twenty tests this exceeds 64\%;
for a hundred it exceeds 99\%. Naive hypothesis testing does
not scale.

Rather than controlling the probability of any false positive (the
family-wise error rate), modern approaches control the False
Discovery Rate: the expected proportion of false positives among
reported discoveries~\cite{benjamini1995fdr}.

\begin{definition}[False Discovery Rate]
Let $V$ denote the number of false positives (true nulls incorrectly
rejected) and $R$ the total number of rejections. The FDR is
$\mathrm{FDR} = \mathbb{E}[V / \max(R, 1)]$.
\end{definition}

Controlling FDR at level $\alpha$ means that on average at most an
$\alpha$ fraction of reported discoveries are false. This is less
conservative than controlling family-wise error rate, preserving
statistical power while maintaining meaningful guarantees.

\subsection{From Batch to Online Testing}

Classical FDR methods like Benjamini--Hochberg~\cite{benjamini1995fdr}
are batch procedures: they require all p-values before making any
decisions. This is unsuitable for AI-Scientists, where hypotheses
arrive sequentially, decisions must be immediate, and experiments may
fail partway through. Online FDR control makes irrevocable
accept/reject decisions as each p-value arrives, keeping the FDR
controlled at every point of a (potentially infinite) sequence.

\subsection{LORD: Alpha-Wealth and Adaptive Thresholds}
\label{sec:lord}

The alpha-wealth framework, introduced by Foster and
Stine~\cite{foster2008alpha} and refined by Aharoni and
Rosset~\cite{aharoni2014generalized}, treats the significance level
as a budget. The procedure maintains a wealth $W(t) \geq 0$; to
test hypothesis $t$, it chooses a threshold $\alpha_t \leq W(t)$,
rejects if $p_t \leq \alpha_t$, and earns a reward upon rejection.
The wealth constraint prevents overspending: the procedure
automatically becomes conservative when discoveries are scarce.

LORD (Levels based On Recent Discovery), introduced by Javanmard and
Montanari~\cite{javanmard2018online}, instantiates this framework with
thresholds that depend on when past discoveries occurred. Let
$\tau_1, \tau_2, \ldots$ denote the discovery times. We use their
version~2 of LORD, with reward $b_0 = \alpha - W_0$ per discovery:
\begin{equation}
\alpha_t \;=\; \gamma_t \cdot W_0 \;+\; (\alpha - W_0)
  \sum_{j:\, \tau_j < t} \gamma_{t - \tau_j}
\label{eq:lord}
\end{equation}
where $\alpha$ is the target FDR level, $W_0 \in (0, \alpha)$ is the
initial wealth, and $\{\gamma_j\}_{j=1}^{\infty}$ is a non-negative
sequence summing to~1 that controls how wealth is distributed over
time. The first term spends initial wealth, decaying via $\gamma_t$.
The second redistributes wealth from past discoveries: each rejection
at time $\tau_j$ contributes $(\alpha - W_0)\cdot\gamma_{t-\tau_j}$
to the current threshold. Recent discoveries contribute more, so
the procedure allocates larger thresholds following consecutive
rejections and smaller thresholds following consecutive
non-rejections. LORD++~\cite{ramdas2017lord} improves on this rule
by paying $\alpha - W_0$ only for the first discovery and $\alpha$ for
every later one; our rule is the more conservative of the two.

A standard choice~\cite{javanmard2018online} is
$\gamma_j = c \cdot \log(\max(j,2)) / (j \cdot e^{\sqrt{\log j}})$
with normalizing constant $c = 0.07720838$. This decays
polynomially, balancing responsiveness with stability.

\paragraph{The FDR guarantee}
A valid test makes its null p-values (super-)uniform: $\mu(P\le x)\le
x$ for all $x\in[0,1]$. Under that assumption, and with the null
p-values independent of the other p-values, LORD controls
$\mathrm{FDR} \leq \alpha$ at every fixed time
(Theorem~3.1 of~\cite{javanmard2018online}). The Lean proof of
Section~\ref{sec:lean} rests on four conditions: the
p-values must be uniform under the null; the thresholds must be
predictable (determined by past information only); future p-values
must be independent of past decisions; and the total wealth spent
must remain bounded. The mathematical core is showing that a certain
conditional expectation equals one almost everywhere (i.e.\ with
probability one), from which the FDR bound follows by the tower law
(the identity $\mathbb{E}[\mathbb{E}[X\mid\mathcal{F}]]=\mathbb{E}[X]$).

\section{A Running Example}
\label{sec:example}

We ground the architecture in a concrete scenario used throughout
the paper: an AI-Scientist optimizing SVM classifiers on the Wine
dataset~\cite{wine_dataset}.

The system starts with a baseline model (a linear SVM with weak
regularization, $C = 0.01$) and five candidate hypotheses for
improvement: (H1)~optimizing RBF kernel parameters via grid search,
(H2)~applying feature scaling, (H3)~tuning the regularization
parameter~$C$, (H4)~switching to polynomial kernels, and
(H5)~performing feature selection via recursive elimination. For each
hypothesis the AI-Scientist generates Python code implementing the
optimization, runs experiments comparing the result to the baseline,
and determines whether the improvement is statistically significant.
The goal is to control FDR at $\alpha = 0.05$.

\paragraph{What goes wrong without safeguards}
Two failures are possible. First, a naive approach tests each
hypothesis at fixed threshold $\alpha = 0.05$. Consider the p-values
in Table~\ref{tab:example_trace}: the naive approach reports three
discoveries (H1, H2, H3). But with five independent tests, the
probability of at least one false positive is already
$1-(1-0.05)^5 \approx 23\%$, and at scale---our Monte Carlo
simulation with two thousand hypotheses
(Section~\ref{sec:simulation})---empirical FDR reaches 41\%.

Second, the LLM-generated code might inadvertently use validation
data during optimization. If \texttt{optimize} loads both
exploration and validation sets, the trained model has seen the
test data, making any subsequent p-value meaningless. No amount of
FDR correction can rescue a biased p-value.

\begin{table}[t]
\centering
\caption{Illustrative LORD trace ($W_0 = 0.1\alpha$) showing how the adaptive threshold
$\alpha_t$ processes a p-value stream: H2's p-value (0.048) would pass
the naive $\alpha = 0.05$ but fails the LORD threshold of 0.00247.
The p-values here are representative inputs chosen to exhibit the
mechanism; the empirical case-study results are in
Table~\ref{tab:casestudy_contrast}.}
\label{tab:example_trace}
\small
\begin{tabular}{@{}clcccc@{}}
\toprule
$t$ & Hypothesis & P-Value & $\alpha_t$ & $p_t \leq \alpha_t$? & Decision \\
\midrule
1 & RBF Kernel Optimization & 0.00009 & 0.00027 & Yes & \textbf{Discovery} \\
2 & Feature Scaling & 0.04784 & 0.00247 & No & Not significant \\
3 & C Regularization & 0.00001 & 0.00057 & Yes & \textbf{Discovery} \\
4 & Polynomial Kernel & 0.32467 & 0.00290 & No & Not significant \\
5 & Feature Selection & 0.13846 & 0.00093 & No & Not significant \\
\bottomrule
\end{tabular}
\end{table}

\paragraph{Two layers of enforcement}
Our architecture addresses both failures through complementary
mechanisms. At the macro level, the \texttt{Research} monad enforces
the LORD protocol: H2's comparison uses the protocol-computed
threshold, and the monad makes it impossible to bypass this check. At
the micro level, a generated Python harness controls data flow so that
the LLM's \texttt{optimize} function receives only exploration
data, and the validation file is physically inaccessible during
optimization. We develop these mechanisms in the next two sections.

\section{The Research Monad: Macro-Level Enforcement}
\label{sec:monad}

This section builds the macro-level defense against the first failure
of Section~\ref{sec:example}---FDR inflation from naive thresholding.
The \texttt{Research} monad provides what we call inevitable
accounting: it is impossible to test a hypothesis without correctly
updating the statistical state.

\subsection{Abstracting the Protocol}

We abstract sequential statistical protocols using a Haskell type
class. An instance must provide three operations: initialization from
a configuration, a pure state-advance function that takes a p-value
and returns a rejection decision along with the updated state, and a
transition validator that checks protocol-specific invariants.
Crucially, \texttt{isValidTransition} has no default implementation,
forcing each protocol to explicitly define its constraints. For
LORD, this enforces strict sequential time advancement---the new
timestamp must equal the old timestamp plus one, catching off-by-one
errors that would silently corrupt the wealth accounting.

\begin{lstlisting}[
    caption={The \texttt{StatisticalProtocol} type class. Each protocol
             must explicitly define its transition constraints.},
    label=lst:typeclass,
    float=tp
]
class StatisticalProtocol s where
    type Config s
    initializeState
      :: Config s -> Either ProtocolError s
    advanceState
      :: Double -> s -> (Bool, Double, s)
    isValidTransition
      :: s -> s -> Either ProtocolError ()
\end{lstlisting}

\subsection{The Monadic Architecture}

The \texttt{Research} monad is a transformer stack combining
\texttt{ExceptT} (for protocol violations) over \texttt{StateT}
(for sequential state threading) over an arbitrary base monad~$m$:

\begin{lstlisting}[
    caption={The \texttt{Research} monad: exception handling over
             stateful sequential computation.},
    label=lst:monad,
    numbers=none
]
type Research s m a =
  ExceptT ResearchError (StateT s m) a
\end{lstlisting}

The ordering of the two layers matters. \texttt{StateT} ensures that the protocol
state (the wealth, discovery times, and current timestamp) is
threaded purely and sequentially through every operation. No hypothesis
test can access stale state or skip an update. \texttt{ExceptT} sits
on top so that if a protocol violation is detected (say, a
non-sequential timestamp), the computation short-circuits immediately,
preserving the integrity of the last valid state.

\subsection{The \texttt{testHypothesis} Operation}

The central operation is a single function, \texttt{testHypothesis},
that orchestrates the complete lifecycle of a hypothesis test. It
retrieves the current state, crosses the IO boundary to execute the
experiment (via the scaffold, described in the next section), and
upon receiving a p-value performs two mandatory steps: it advances the
protocol state via the pure \texttt{advanceState} function, then
validates the transition. If validation fails, the monad halts with a
structured error. If it succeeds, the new state is committed. The
function is polymorphic in both the protocol type~\texttt{s} and the
base monad~\texttt{m}, so the same orchestration logic works with any
protocol instance.

The key property is that these steps are inescapable. A
programmer using the \texttt{Research} monad cannot test a hypothesis
and forget to update the state---the type system requires it. Nor can
they silently swallow a validation error---\texttt{ExceptT} propagates
it. The full implementation is available in the repository
(\texttt{ResearchMonad.hs}); we omit it here for brevity, since the
structure is standard for monad transformer stacks and the essential
design is captured by the type signature and the protocol abstraction
above.

\paragraph{Running example: FDR accounting}
Table~\ref{tab:example_trace} shows the monad in action on the SVM
scenario. H2's p-value of 0.048 would pass the naive $\alpha = 0.05$
but fails the LORD threshold of $\alpha_2 = 0.00247$. The crucial
point is that no code path exists to compare H2's p-value
against 0.05 instead: \texttt{advanceState} computes the threshold
from the protocol state, and the rejection decision is made inside
this pure function. The first failure from
Section~\ref{sec:example}---FDR inflation through naive
thresholding---is eliminated by construction.

\section{Declarative Scaffolding: Micro-Level Enforcement}
\label{sec:scaffolding}

The \texttt{Research} monad guarantees correct FDR accounting but
relies on unbiased p-values as input. When \texttt{testHypothesis}
crosses the IO boundary into LLM-generated Python code, we enter
untrusted territory. Relying on prompts to make the LLM generate
methodologically sound code (``please do not use validation data
during training'') is brittle. We instead employ structural
enforcement: the architecture makes certain errors impossible rather
than asking the LLM to avoid them.

The mechanism is straightforward. The Haskell orchestrator generates a Python
harness---a rigid skeleton that controls data flow and fixes the
statistical test---before the LLM writes any code. The LLM then fills
in domain-specific logic (the \texttt{optimize} and
\texttt{evaluate\_model} functions) within the constraints the
harness imposes (Listing~\ref{lst:harness}).

\begin{lstlisting}[
    caption={Generated Python harness. The LLM's \texttt{optimize}
             function receives only exploration data; validation data
             and the statistical test are controlled by the harness.},
    label=lst:harness,
    float=tp,
    style=PythonClean
]
# harness.py (Generated by Haskell Orchestrator)
import implementation  # LLM-generated code
import pandas as pd
from verified_stats import paired_permutation_pvalue

def run_exploration():
    data = pd.read_csv("data/wine_exploration.csv")
    artifact = implementation.optimize(data.copy())
    return artifact

def run_validation(artifact, baseline, hyp_id):
    # disjoint, pre-assigned split for this
    # hypothesis (H3)
    data = pd.read_csv(
        f"data/wine_validation_H{hyp_id}.csv")
    # per-example losses, computed by the harness
    loss_a = per_example_loss(artifact, data)
    loss_b = per_example_loss(baseline, data)
    # paired sign-flip permutation test:
    # super-uniform nulls (H1)
    return paired_permutation_pvalue(loss_a, loss_b)
\end{lstlisting}

The harness enforces three properties structurally. First, data
separation: the LLM's code receives only the exploration data path;
each hypothesis's disjoint validation split is loaded by the harness
during validation, not by the LLM during optimization. Second, test
standardization: the statistical test---a paired permutation test whose
null p-values are super-uniform by construction---is executed by the
harness on per-example losses, not by LLM-generated code. Third,
interface constraints: the LLM must
implement a fixed API (\texttt{optimize}, \texttt{get\_baseline},
\texttt{evaluate\_model}), limiting the scope for methodological
errors. These contracts are defined as Haskell types
(\texttt{DataContract} and \texttt{StatisticalTestSpec}) and are
available in the repository.

\subsection{OS-Level Data Separation}
\label{sec:sandbox}

Passing only the exploration split to \texttt{optimize} is an
interface-level restriction: it constrains the arguments the function
receives, but not the files the process can open. LLM-generated code
runs with the same file-system access as the interpreter itself, so a
single \texttt{pd.read\_csv} with a guessed or hard-coded path---or a
subprocess it launches---could read the validation file directly,
defeating the contract. A robust guarantee must remove the validation data from the
execution environment itself, not merely from the function signature.

We obtain this with an operating-system boundary, which we implement
and test as a standalone component. It executes the exploration phase
inside an unprivileged Linux
\emph{user}+\emph{mount} namespace, created with the standard Linux
\texttt{unshare} facility; no root, container runtime, or external
dependency is required. Within the namespace, the data directory is
replaced by a view that contains only the exploration split, so the
validation file is \emph{absent} from the filesystem the optimization
code observes. Because this view is inherited by every child process it
starts, the restriction cannot be escaped by an absolute path, by
scanning the filesystem, or by launching a helper process: there is no
path, in any process under the optimization, that resolves to the
validation data. The validation
phase runs afterwards, in a separate step, on the already-frozen
artifact.

This converts ``the code is asked not to read the validation data''
into ``the validation data does not exist where the code runs.'' The
guarantee holds for harness-private data---data that exists only as the
files the orchestrator controls. It does not extend to data the model
can obtain by other means; in particular, a public benchmark that
ships inside a library (such as the Wine dataset, available through
\texttt{sklearn.datasets}) can be reconstructed without touching any
file, which no filesystem boundary can prevent. We evaluate the
guarantee, and scan generated code for this separate channel, in
Section~\ref{sec:leakage}. In the current release the orchestrator
does not yet invoke this component: the case studies of
Section~\ref{sec:casestudy} rely on the interface-level separation of
the harness, and the sandbox is exercised directly by the adversarial
evaluation.

\paragraph{Running example: data separation}
Recall the second failure from Section~\ref{sec:example}: an LLM
might load validation data during optimization, producing artificially
low p-values. Consider H1 (RBF kernel optimization). When
\texttt{run\_explora\-tion} runs inside the namespace of
Section~\ref{sec:sandbox}, the data directory exposes only
\texttt{wine\_explora\-tion.csv}; the trained artifact is then frozen
and handed to \texttt{run\_valida\-tion}, executed separately. Even if
the LLM's code contains \texttt{pd.read\_csv("wine\_valida\-tion\_H1.csv")}
with the correct absolute path, the open fails: the file is not present
in the optimization's mount namespace, and a subprocess it spawns
inherits the same view. The statistical test
(\texttt{paired\_permutation\_pvalue}, a paired sign-flip permutation
test on held-out per-example losses) is called by the harness, not by
the LLM, so the LLM cannot substitute a weaker test or manipulate its
parameters. The
second failure (data leakage invalidating p-values) is eliminated
structurally for harness-private data. Together with the monad's
protection against FDR inflation, the two layers close both gaps
identified in Section~\ref{sec:example}.

\section{Formal Foundations: LORD in Lean~4}
\label{sec:lean}

The architecture described above enforces the LORD protocol. The
FDR guarantee itself rests on the original measure-theoretic proof by
Javanmard and Montanari~\cite{javanmard2018online}. But citing that
proof does not say which of its assumptions an implementation must
satisfy, nor in what form.

The formalization we present addresses this gap. By encoding the
theorem in Lean~4's dependent type system, we obtain the four
sufficient conditions for FDR control as machine-readable type
signatures. These type signatures serve as a specification contract:
a precise statement of what an implementation must satisfy, against
which the Haskell architecture can be compared term by term.

\subsection{What We Formalize}

The formalization spans roughly $1{,}450$ lines in five files, plus a
213-line file of basic definitions, with zero uses of
\texttt{sorry} (Lean's escape hatch for unproved claims). Two files
form its measure-theoretic core, described next; three further files
derive the budget and the marginal and full FDR theorems
(Section~\ref{sec:faithfulness}).

The first file, \texttt{FundamentalLemma.lean} (660~lines),
establishes the measure-theoretic core. We define a uniform p-value
as a random variable $P$ satisfying $\mu(P \leq t) = t$ for all
$t \in [0,1]$, and define independence from a sub-$\sigma$-algebra
$\mathcal{F}_t$ as the condition that $P^{-1}(B)$ is independent of
every $\mathcal{F}_t$-measurable set. The central result is:
\[
\mathbb{E}\!\left[\frac{\mathbf{1}\{P \leq \alpha\}}{\alpha}
  \;\Big|\; \mathcal{F}_t\right] = 1 \quad \text{a.e.}
\]
This ``fundamental lemma'' says that the indicator-over-threshold ratio
has conditional expectation one---the property that drives the entire
FDR proof. The Lean proof proceeds by first establishing the
unconditional set-integral identity
$\int_A \mathbf{1}\{P \leq \alpha\}/\alpha \, d\mu = \mu(A)$ for
every $\mathcal{F}_t$-measurable set $A$, which characterizes the
conditional expectation by its defining property.

The second file, \texttt{OnlineFDR.lean} (195~lines), assembles the
FDR bound from the fundamental lemma. Its core identity is that the
expected number of false discoveries equals the expected total
threshold spent on the nulls, $\mathbb{E}[V] = \sum_{t \in H_0} \mathbb{E}[\alpha_t]$;
the budget then bounds that sum, and since $V/\max(R,1) \leq V$, the
false discovery rate inherits the bound:
\[
\mathrm{FDR} = \mathbb{E}\!\left[\frac{V}{\max(R,1)}\right] \leq q.
\]

\subsection{The Theorem Statement and Its Faithfulness}
\label{sec:faithfulness}

A machine-checked proof guarantees only that the \emph{stated} theorem
holds; it says nothing about whether that statement is the one the
reader has in mind. To let the reader judge faithfulness directly, we
reproduce the relevant definitions and the top-level theorem verbatim
from \texttt{OnlineFDR.lean}, then relate each component to its
textbook counterpart and state precisely what is and is not covered.

The three definitions that carry the semantic weight are shown in
Listing~\ref{lst:leandefs}.
\begin{lstlisting}[
    style=LeanClean,
    caption={Definitions underlying the theorem (verbatim from the
             repository; line breaks adjusted). \texttt{Rej}/\texttt{sumRej} build the
             FDR ratio \texttt{V/max(R,1)}.},
    label=lst:leandefs,
    float=tp
]
def IsUniformPValue {Ω : Type*}
    [MeasurableSpace Ω]
    (μ : Measure Ω) (P : Ω → ℝ) : Prop :=
  ∀ x : ℝ, 0 ≤ x → x ≤ 1 →
    μ {ω | P ω ≤ x} = ENNReal.ofReal x

def IsIndepOfSubalgebra {Ω : Type*}
    [MeasurableSpace Ω]
    (μ : Measure Ω) (P : Ω → ℝ)
    (m₀ : MeasurableSpace Ω) : Prop :=
  ∀ (A : Set Ω) (B : Set ℝ),
    @MeasurableSet Ω m₀ A → MeasurableSet B →
    μ (A ∩ P ⁻¹' B) = μ A * μ (P ⁻¹' B)

noncomputable def Rej {Ω : Type*}
    (P α : Ω → ℝ) (ω : Ω) : ℝ :=
  if P ω ≤ α ω then 1 else 0

noncomputable def sumRej {Ω : Type*}
    (P α : ℕ → Ω → ℝ)
    (S : Finset ℕ) (ω : Ω) : ℝ :=
  ∑ t ∈ S, Rej (P t) (α t) ω
\end{lstlisting}
The top-level theorem is shown in Listing~\ref{lst:leanthm}.
\begin{lstlisting}[
    style=LeanClean,
    caption={The top-level theorem \texttt{lord\_fdr}, verbatim
             (line breaks adjusted; comment lines added).
             The conclusion is exactly \texttt{E[V/max(R,1)] <= q}.},
    label=lst:leanthm,
    float=tp
]
theorem lord_fdr
    [IsProbabilityMeasure μ]
    {P α : ℕ → Ω → ℝ} {F : ℕ → MeasurableSpace Ω}
    {ε : ℝ}
    (hε : 0 < ε)
    (hF : ∀ t, F t ≤ mΩ)
    (hPm : ∀ t, @Measurable Ω ℝ mΩ _ (P t))
    (H₀ S : Finset ℕ)
    -- (H1) uniform null p-values
    (hU : ∀ t ∈ H₀, IsUniformPValue μ (P t))
    -- (H2) predictable thresholds
    (hαp : ∀ t, @Measurable Ω ℝ (F t) _ (α (t + 1)))
    (hαlo : ∀ t, ∀ᵐ ω ∂μ, ε ≤ α t ω)
    (hαhi : ∀ t, ∀ᵐ ω ∂μ, α t ω ≤ 1)
    -- (H3) nulls independent of the past
    (hI : ∀ t, t + 1 ∈ H₀ →
      IsIndepOfSubalgebra μ (P (t + 1)) (F t))
    (hH₀ : ∀ t ∈ H₀, ∃ s, t = s + 1)
    (hS : ∀ t ∈ S, ∃ s, t = s + 1)
    (hH₀S : H₀ ⊆ S)
    -- (H4) budget
    {q : ℝ} (hbudget : ∀ᵐ ω ∂μ, ∑ t ∈ S, α t ω ≤ q) :
    ∫ ω, sumRej P α H₀ ω / max (sumRej P α S ω) 1 ∂μ
      ≤ q
\end{lstlisting}
Reading the statement against the literature confirms that no
definition has been weakened to ease the proof:

\smallskip
\noindent
\textbf{The conclusion is the textbook FDR.} \texttt{sumRej P}
\texttt{$\alpha$ S} is $R = \sum_{t \in S} \mathbf 1\{P_t \le
\alpha_t\}$, the rejection count; \texttt{sumRej P $\alpha$ H$_0$} with
\texttt{H$_0$ $\subseteq$ S} is $V$, the rejections among the true-null
indices. The integrand $V/\max(R,1)$ and the conclusion
$\int \cdots \le q$ are therefore exactly
$\mathbb{E}[V/\max(R,1)] = \mathrm{FDR} \le q$ of
Definition~1---not a surrogate such as $\mathbb{E}[V]/\mathbb{E}[R]$.

\smallskip
\noindent
\textbf{The hypotheses are the four conditions, unmodified.}
\texttt{hU} is uniformity of each null p-value; \texttt{hI} is the
measure-theoretic independence of each null $P_{t+1}$ from the past
$\sigma$-algebra $F_t$; \texttt{h$\alpha$p} is
$F_t$-measurability (predictability) of $\alpha_{t+1}$; and
\texttt{hbudget} is the budget constraint
$\sum_{t \in S}\alpha_t \le q$. These correspond to the assumptions
of the classical results~\cite{javanmard2018online,ramdas2017lord}:
uniform and independent null p-values and predictable thresholds
appear there directly, and (H4) is our reformulation of the wealth
constraint, which Javanmard and Montanari state as $w_0 + b_0 \le
\alpha$ on a generalized alpha-investing rule.

\smallskip
\noindent
\textbf{Scope of the claim.} Four points keep the
claim precise. (i)~\emph{Fixed horizon.} The theorem quantifies over
arbitrary finite index sets \texttt{H$_0$ $\subseteq$ S}~(\texttt{Finset
$\mathbb{N}$}), so it establishes FDR control at every fixed
deterministic horizon, which is the form an implementation running a
bounded experiment campaign requires; it is not the optional-stopping
formulation, whose extension we leave to future work.
(ii)~\emph{The budget is derived, not assumed.} Although
\texttt{hbudget} appears as a hypothesis of \texttt{lord\_fdr}, the
budget is not simply postulated: a separate machine-checked lemma
(\texttt{lordThreshold\_sum\_le}) proves that the LORD thresholds
(Equation~\ref{eq:lord}) satisfy the pathwise bound
$\sum_{t} \alpha_t \le \alpha\cdot\max(R,1)$ from $\sum_j\gamma_j\le 1$
alone, with the reward term active. Because each discovery refunds
wealth, the cumulative spend can exceed $W_0$; the correct budget is
therefore this $\alpha\cdot\max(R,1)$ bound, \emph{not} a pathwise
$\sum_t\alpha_t\le W_0$. The remaining floating-point obligation is
that the computed thresholds never exceed the computed wealth, which
Section~\ref{sec:spark-eq1} proves over IEEE~754 under a margin
condition. (iii)~\emph{Exact versus super-uniform.}
\texttt{IsUniformPValue} states equality $\mu\{P\le x\}=x$; the
classical result also holds under super-uniformity
($\mu\{P\le x\}\le x$). Assuming equality is therefore mildly
stronger (it narrows applicability, never the converse); we have not
carried out the relaxation, which touches the fundamental lemma, where
uniformity enters as an equality. (iv)~\emph{Thresholds bounded away
from zero.} \texttt{h$\alpha$lo} requires $\varepsilon \le \alpha_t$
for every $t$. LORD thresholds satisfy $\alpha_t \ge \gamma_t W_0 > 0$,
so on a finite horizon a positive $\varepsilon$ exists; beyond the
horizon the thresholds can be set to any constant in $[\varepsilon,
1]$ without affecting the conclusion, which concerns only
$S$. Time indices start at~1 (\texttt{hH$_0$}, \texttt{hS}). None of
these caveats weakens the conclusion; they delimit the conditions under
which it is proved.

\paragraph{Derived budget, marginal FDR and full FDR}
Three further machine-checked results sharpen the statement above (all
\texttt{sorry}-free, depending only on Lean's three standard axioms).
First, the derived budget of point~(ii)
(\texttt{lordThreshold\_sum\_le}) replaces the assumed constant budget
with a bound proved from the LORD threshold formula. Second, \texttt{lord\_mfdr} establishes marginal FDR control,
$\mathbb{E}[V]\le\alpha\,\mathbb{E}[\max(R,1)]$, for the
\emph{reward-bearing} procedure the implementation actually
runs---the adaptive regime in which the cumulative spend is not
pathwise bounded by $W_0$. Third, \texttt{fdr\_le} establishes the
full $\mathrm{FDR}=\mathbb{E}[V/\max(R,1)]\le\alpha$ by a
leave-one-out argument, under the additional condition that each null
p-value is independent of the leave-one-out information and the
thresholds are non-adaptive with respect to it---precisely the regime
that the disjoint-split scaffold of Section~\ref{sec:scaffolding}
creates. Full FDR control for the adaptive reward dynamics, where this
condition fails, remains the pen-and-paper result of Javanmard and
Montanari (Theorem~3.1 of~\cite{javanmard2018online}), now applicable
because the scaffold supplies the independence it requires.

\subsection{The Specification Contract}
\label{sec:bridge}

The Lean theorem \texttt{lord\_fdr} requires four hypotheses, which we
can read directly from its type signature. Each corresponds to a
concrete obligation that an implementation must discharge:

\smallskip
\noindent
\textbf{(H1) Uniform p-values} (\texttt{hU}). Each null p-value must
satisfy $\mu(P_t \leq x) \leq x$. This is a property of the
\emph{test}, not merely of who computes it: fixing a trusted
implementation removes the LLM from the statistical computation but
does not by itself make the nulls uniform. A cross-validated paired
$t$-test in particular treats overlapping folds as independent
observations and is anti-conservative under the
null~\cite{dietterich1998approximate,bengio2004nounbiased}. The
scaffold therefore evaluates each hypothesis with a paired sign-flip
permutation test on \emph{independent held-out per-example losses},
computed by the harness; under the exchangeability null its p-values
are super-uniform in finite samples and distribution-free, which is
condition (H1) in its super-uniform form. The Lean definition
\texttt{IsUniformPValue} states the equality version, a mildly stronger
condition (point~(iii) of Section~\ref{sec:faithfulness}). A null-calibration
experiment (Section~\ref{sec:simulation}) confirms this empirically:
the permutation test holds its nominal level while the cross-validated
$t$-test inflates it.

\smallskip
\noindent
\textbf{(H2) Predictability} (\texttt{h$\alpha$p}). The threshold
$\alpha_{t+1}$ must be $\mathcal{F}_t$-measurable: it can depend only
on information available at time~$t$. This is the condition most
naturally enforced by the monadic architecture. In the
\texttt{Research} monad, the state at step~$t$ is exactly the
information available at time~$t$---it contains the wealth, the
discovery times $\tau_1, \ldots, \tau_k$, and the current timestamp,
but not future p-values. The function \texttt{calculateNextAlpha}
(Equation~\ref{eq:lord}) takes only this state as input. This is not a
runtime check; it is a consequence of how \texttt{StateT} threads
information. The monad makes the wrong program---one where
$\alpha_{t+1}$ peeks at $P_{t+1}$---untypeable.

\smallskip
\noindent
\textbf{(H3) Independence} (\texttt{hI}). Future p-values must be
independent of past decisions. The scaffold's data separation enforces
this by assigning each hypothesis a \emph{disjoint} validation split,
fixed before any testing, so its p-value depends only on rows that no
earlier decision examined. This is essential precisely because
hypotheses are generated adaptively---the LLM proposes hypothesis
$t{+}1$ after seeing the outcome of hypothesis $t$---which would
invalidate a single reused held-out set however carefully it is
loaded~\cite{dwork2015reusable}; disjoint splits break this adaptive
dependence by construction. The cost is data: each hypothesis receives
roughly a $1/m$ fraction of the validation pool. Where that is too
expensive, a reusable holdout~\cite{dwork2015reusable} trades added
noise for reuse; we adopt disjoint splits for their exact, assumption-free
guarantee.

\smallskip
\noindent
\textbf{(H4) Budget constraint} (\texttt{hbudget}). The thresholds must
not overspend the error budget. Over the real numbers the budget is
derived in Lean from the threshold formula and $\sum_j \gamma_j \le 1$
(point~(ii) above), so what remains for an implementation is keeping
the wealth process non-negative under floating-point arithmetic.
This is the one condition our type system does not enforce statically:
it involves the $\gamma$-sequence (Equation~\ref{eq:lord}), computed
with transcendental functions, and IEEE~754 rounding, which resist
lightweight verification. Our Monte Carlo simulation
(Section~\ref{sec:simulation}) illustrates the resulting behavior, and
we give machine-checked proofs over IEEE~754 arithmetic, in SPARK
(Sections~\ref{sec:spark} and~\ref{sec:spark-eq1}).

\subsection{Real Arithmetic Versus Floating Point}
\label{sec:gap}

The Lean proof works over the real numbers $\mathbb{R}$; any
implementation computes with IEEE~754 floating-point arithmetic.  This
opens a question that citation of the original theorem cannot answer:
can rounding errors cause the budget constraint to be violated?
More subtly, if the $\gamma$-sequence partial sums exceed~1 in
floating point, the wealth could be overdrawn by an amount invisible
to real-arithmetic analysis.  Formally connecting floating-point
computation to real-analysis proofs is a known open problem in
verified scientific computing, addressed for example by the Flocq
library~\cite{boldo2011flocq}; verified compilers such as
CompCert~\cite{leroy2009compcert} address the neighboring question of
preserving program meaning through compilation.

Of the four conditions, only the budget constraint (H4) touches this
gap. The structural conditions (H1--H3) concern which functions run and
what data they may see, settled by the type system and scaffold over
$\mathbb{R}$ and floating point alike; H4 instead requires that the
wealth process remain non-negative under a specific sequence of
IEEE~754 multiplications and additions. Types cannot settle that;
verified numerics can. We address it using
SPARK~\cite{mccormick2015spark}, a subset of Ada designed for formal verification, as
described next.

\subsection{Verified Floating-Point Arithmetic in SPARK}
\label{sec:spark}

The arithmetic obligation (H4) is that the wealth process remains
non-negative, $W(t) \geq 0$ for all~$t$.  We stress that this is
\emph{not} the same as a pathwise spend bound
$\sum_t \alpha_t \leq W_0$: once a rejection refunds wealth, the
cumulative spend can exceed the initial budget, so the two are not
equivalent under the reward dynamics.  What the floating-point
implementation must guarantee is the non-negativity invariant itself,
on which the well-definedness of the thresholds
$\alpha_t = \gamma_t W(t)$ rests.  An implementation
must ensure that this invariant is preserved not over $\mathbb{R}$ but
over IEEE~754 double-precision arithmetic, where each operation
introduces rounding error.

To verify this, we re-implement the core wealth update in
SPARK/\allowbreak Ada~2012, a language designed for high-assurance software in
which pre- and postconditions are statically discharged by the
GNATprove toolchain.  The SPARK code covers only the arithmetic on
which the budget depends, for a multiplicative alpha-investing wealth
process---spend
$\alpha_t = \gamma_t W(t)$, earn $\alpha - W_0$ on each
rejection---while the Monte Carlo driver and gamma-sequence
computation remain in ordinary Ada. This process shares LORD's wealth
accounting but not its threshold formula (Equation~\ref{eq:lord});
Section~\ref{sec:spark-eq1} verifies the thresholds of
Equation~\ref{eq:lord} themselves.

The key design decision is to express the wealth update as a
multiplication rather than a subtraction
(Listing~\ref{lst:spark}).
\begin{lstlisting}[
    style=AdaClean,
    caption={SPARK wealth update (excerpt). GNATprove
             verifies that \texttt{Wealth >= 0.0} holds
             after every step.},
    label=lst:spark,
    float=tp
]
  One_Minus_G := 1.0 - Gamma_T;
  pragma Assert (One_Minus_G >= 0.0);

  Alpha_T  := Gamma_T * S.Wealth;
  S.Wealth := One_Minus_G * S.Wealth;
  pragma Assert (S.Wealth >= 0.0);  -- H4

  if Reject then
    Reward   := S.Alpha_Param - S.W0;
    S.Wealth := S.Wealth + Reward;
    pragma Assert (S.Wealth >= 0.0);
  end if;
\end{lstlisting}
The formulation $W_{t+1} = (1 - \gamma_t) \times W_t$ is
mathematically equivalent to $W_t - \gamma_t W_t$, and both are in
fact non-negative under IEEE~754: rounding is monotone, so the computed
product $\gamma_t W_t$ never exceeds $W_t$ when $\gamma_t \in [0,1]$,
and the difference of two floating-point numbers $a \geq b$ is never
negative. The two forms differ in what the prover can show. The
product of two non-negative finite values is non-negative, and
GNATprove discharges the multiplicative form automatically; for the
subtraction it would need the monotonicity of rounding, and at proof
levels~2 and~4 it leaves the corresponding range check unproved. The
multiplicative form is what lets the proof go through without manual
lemmas.

The gamma-sequence computation involves transcendental functions
(\texttt{log}, \texttt{exp}, \texttt{sqrt}) that are outside SPARK's
proof scope.  We bridge this gap with a SPARK-verified clamping
function whose postcondition \texttt{Result in 0.0..1.0} is
statically discharged, ensuring the precondition of the wealth update
is satisfied regardless of floating-point rounding in the gamma
formula:
\begin{lstlisting}[
    style=AdaClean,
    caption={SPARK clamping function ensuring output stays in valid probability range.},
    label=lst:clamp
]
function Safe_Gamma (Raw : Long_Float)
   return Probability
  with Post => Safe_Gamma'Result in 0.0 .. 1.0
is
begin
   if Raw <= 0.0 then return 0.0;
   elsif Raw >= 1.0 then return 1.0;
   else return Raw;
   end if;
end Safe_Gamma;
\end{lstlisting}

Beyond the per-step invariant, we verify the full inductive closure
across an arbitrary hypothesis sequence using a SPARK procedure with
an explicit loop invariant (Listing~\ref{lst:loop}).
\begin{lstlisting}[
    style=AdaClean,
    caption={SPARK loop invariant establishes full inductive closure:
             \texttt{Wealth >= 0.0} holds throughout the sequence.},
    label=lst:loop,
    float=tp
]
procedure Run_Sequence
  (S : in out Protocol_State; ...)
  with Post => S.Wealth >= 0.0
is
begin
   for I in Gammas'Range loop
      pragma Loop_Invariant (S.Wealth >= 0.0);
      pragma Loop_Invariant (S.W0 > 0.0);
      pragma Loop_Invariant (S.W0 < S.Alpha_Param);
      pragma Loop_Invariant (S.Alpha_Param <= 1.0);
      Advance (S, Gammas(I), P_Values(I),
               Alpha_T, Rejects(I));
   end loop;
end Run_Sequence;
\end{lstlisting}
GNATprove verifies that the loop invariant is established on entry
and preserved by each iteration, using the postconditions of
\texttt{Advance} which guarantee that \texttt{S.Wealth >= 0.0} and
that the constants \texttt{S.W0} and \texttt{S.Alpha\_Param} are
unchanged.

The SPARK contract captures the inductive step of the budget
invariant:
\begin{quote}
\textbf{Precondition.}\;
$\mathtt{Wealth} \geq 0.0
  \;\wedge\; \gamma_t \in [0, 1]
  \;\wedge\; W_0 \in (0, \alpha)
  \;\wedge\; \alpha \in (0, 1]$

\smallskip
\textbf{Postcondition.}\;
$\mathtt{Wealth} \geq 0.0
  \;\wedge\; \alpha_t \geq 0.0
  \;\wedge\; W_0' = W_0
  \;\wedge\; \alpha' = \alpha$
\end{quote}
The postcondition includes preservation of the constants $W_0$ and
$\alpha$, which GNATprove uses to verify the loop invariants in
\texttt{Run\_Sequence}.

GNATprove (version~15.1.0, proof level~2, a fixed budget of
100{,}000 prover steps) discharges all 30 verification conditions of
the protocol package \texttt{Lord\_PP}; the C interface to the kernel
(\texttt{Lord\_Capi}, Section~\ref{sec:division}) adds five more, also
discharged. \texttt{Lord\_PP} needs no manual lemmas, and its proofs
finish within the default time limit of proof level~2. The base case
is immediate: \texttt{Initialize} sets $W := W_0 > 0$.

\subsection{The Thresholds of Equation~(\ref{eq:lord}) in SPARK}
\label{sec:spark-eq1}

The package \texttt{Lord\_PP} verifies a multiplicative wealth process,
not the thresholds of Equation~\ref{eq:lord} that the Lean theorems and
the Haskell implementation use. A second package,
\texttt{Lord\_Eq1}, computes Equation~\ref{eq:lord} itself in
\texttt{Long\_Float}, adding the sum over past discoveries left to
right, and maintains the generalized alpha-investing wealth
$W(t) = W(t-1) - \alpha_t + (\alpha - W_0) R_t$, $W(0) = W_0$. Over
$\mathbb{R}$ this rule never spends more than its wealth,
$\alpha_t \le W(t-1)$ (Lemma~2.4 of~\cite{javanmard2018online}). The
question is whether the same holds for the thresholds computed in
floating point.

\paragraph{Without a margin, it does not}
When the prefix sums of $\gamma$ reach~1, rounding can make the
computed threshold exceed the computed wealth by one unit in the last
place. In a two-step example with
$\alpha = \texttt{0x1.f2712dbb241fcp-1}$,
$W_0 = \texttt{0x1.f34a59a90138ep-2}$,
$\gamma_1 = \texttt{0x1.d06019087de8ap-2}$ and
$\gamma_2 = \texttt{0x1.17cff37bc10bbp-1}$, the floating-point sum
$\gamma_1 + \gamma_2$ equals $1.0$, and at step~2 the threshold exceeds
the wealth by $5.55 \times 10^{-17}$. A random search finds such cases
both when the floating-point prefix sums of $\gamma$ are at most~1 and
when their exact real sums are. The budget proved over $\mathbb{R}$
therefore does not carry over to IEEE~754 unchanged.

\paragraph{Two proved results}
The first result needs only $\gamma_t \in [0,1]$. The kernel spends
$\alpha_t = \min(\alpha_t^{\mathrm{Eq}}, W(t-1))$, where
$\alpha_t^{\mathrm{Eq}}$ is the floating-point value of
Equation~\ref{eq:lord}, so $\alpha_t \le W(t-1)$ before every update
and $W(t) \ge 0$ after it. A counter \texttt{Clamp\_Count} records every
step at which the minimum changes the value; at every other step
$\alpha_t = \alpha_t^{\mathrm{Eq}}$. GNATprove also proves the absence of
run-time errors (overflow, range and index checks) and
$\alpha_t \ge 0$.

The second result removes the minimum. Let $P(n)$ be the exact real
sum of the floating-point values $\gamma_1, \ldots, \gamma_n$, and let
$M = 10{,}000 \cdot 10{,}005 \cdot 2^{-38} \approx 3.64 \times 10^{-4}$.
For horizons up to
$10{,}000$ steps, if $W_0 (1 - P(n)) \ge M$ for every
$n \le 10{,}000$,
then $\alpha_t^{\mathrm{Eq}} \le W(t-1)$ at every step:
\texttt{Clamp\_Count} stays~0, and the kernel computes exactly
Equation~\ref{eq:lord}. The proof carries a ghost invariant in exact
real arithmetic (SPARKlib's \texttt{Big\_Real}): after $T$ steps the
floating-point wealth lies at most $T \cdot 10{,}005 \cdot 2^{-38}$
below its real-number value, with each rounding bounded by SPARKlib's
rounding-error lemmas. All \texttt{Big\_Real} code is ghost, so the
compiled kernel is pure \texttt{Long\_Float} arithmetic. The procedures
\texttt{Run\_Sequence} and \texttt{Run\_Sequence\_Exact} carry the two
results across a whole sequence of p-values.

Our configurations meet the margin with room to spare:
$P(10{,}000) = 0.384$, so
$W_0 (1 - P) = 3.08 \times 10^{-3}$ for $W_0 = 0.1\alpha$ and
$1.54 \times 10^{-2}$ for $W_0 = \alpha/2$: 8.5 and 42 times $M$.

\paragraph{Agreement with the implementation}
The verified kernel reproduces the Haskell implementation bit for bit:
every $\alpha_t$ on 200{,}000 steps of the Monte Carlo configuration of
Section~\ref{sec:simulation} and on the
sequences of Tables~\ref{tab:example_trace},
\ref{tab:casestudy_contrast} and~\ref{tab:casestudy_positive}, with no
reject decision differing and \texttt{Clamp\_Count} equal to~0
throughout. A Haskell harness that calls the kernel through a C
interface (\texttt{Lord\_Eq1\_Capi}) reproduces the same values.

\paragraph{Proof effort}
For the whole SPARK project, GNATprove~15.1.0 with SPARKlib~15.1.0
(proof level~2, 100{,}000 prover steps, 60~seconds per check)
discharges 778 checks, 716 by provers and 62 by flow analysis, with none
unproved (Table~\ref{tab:spark_counts}). The unit with the second
result is compiled as Ada~2022, the others as Ada~2012. No proved unit uses
\texttt{pragma Assume}, a justification annotation, or
\texttt{SPARK\_Mode => Off}.

\begin{table}[t]
\centering
\caption{Checks discharged by GNATprove provers, per SPARK unit.}
\label{tab:spark_counts}
\small
\begin{tabular}{@{}llr@{}}
\toprule
Unit & Content & Prover checks \\
\midrule
\texttt{Lord\_PP} & Multiplicative wealth process & 30 \\
\texttt{Lord\_Capi} & C interface to \texttt{Lord\_PP} & 5 \\
\texttt{Lord\_Eq1} & Equation~\ref{eq:lord} with minimum & 138 \\
\texttt{Lord\_Eq1\_Exact} & Equation~\ref{eq:lord} without minimum, with margin & 532 \\
\texttt{Lord\_Eq1\_Capi} & C interface to \texttt{Lord\_Eq1} & 11 \\
\midrule
Total & (plus 62 flow checks) & 716 \\
\bottomrule
\end{tabular}
\end{table}

\subsection{What the SPARK Proofs Cover}
\label{sec:spark-scope}

The SPARK proofs directly address H4 and, as a side effect,
independently confirm H2.  The structural conditions H1 and H3 are
addressed exactly as described in Section~\ref{sec:bridge}---by the
harness-owned test and the disjoint validation splits of the
scaffold---and SPARK adds nothing to
their enforcement.  For H2 and H4, however, SPARK provides
independent evidence.

For H2 (predictability), GNATprove's flow analysis verifies that
\texttt{Advance} reads only from the current protocol
state~\texttt{S}, the caller-supplied \texttt{Gamma\_T}, and the
current \texttt{P\_Value}, with no access to future state or global
variables.  This is not merely a consequence of good coding practice;
it is a machine-checked data-flow property that independently
confirms the \texttt{StateT}-based enforcement of
Section~\ref{sec:bridge}.

H4 (budget constraint) is the central contribution of the SPARK
proofs. For the multiplicative process, no sequence of IEEE~754
rounding operations can make the wealth negative. For the thresholds
of Equation~\ref{eq:lord}, the computed threshold never exceeds the
computed wealth: always for the kernel with the minimum, and under the
margin condition without it. The Lean theorems establish the budget
over $\mathbb{R}$ for the same formula. The two proofs therefore concern
the same thresholds in different arithmetic, but no machine-checked
refinement connects them.

\paragraph{Remaining limitations}
The Lean and SPARK proofs are linked by the shared threshold formula,
not by a machine-checked refinement from real to floating-point
arithmetic. The $\gamma$ table is computed outside SPARK with
\texttt{log}, \texttt{exp} and \texttt{sqrt}; the SPARK proofs take it
as an input, and the margin condition is a precondition on the table
that is not checked at run time. Our experiments run on the Haskell
implementation, which the verified kernel matches bit for bit on the
configurations above but does not replace.

\subsection{Traceability}
\label{sec:traceability}

Table~\ref{tab:conditions} collects the links established in
Sections~\ref{sec:faithfulness}--\ref{sec:spark-scope}, from the
top-level requirement to its enforcement, into a traceability matrix.
Each row traces one requirement to its formal statement in
Lean, the mechanism that enforces it, and the evidence that it holds.
The status column separates three kinds of link: \emph{machine-checked}
(a Lean or SPARK proof), \emph{enforced by construction} (a program that
violates the requirement does not typecheck or never receives the data),
and \emph{assumed} (a requirement the architecture supports but does not
prove, backed by experiment). (H1) is the only assumed requirement.

\begin{table}[!htbp]
\centering
\caption{Traceability matrix: the top-level requirement and the four
derived requirements (H1)--(H4), each traced to its Lean statement, its
enforcement mechanism, the evidence that it holds, and its status.}
\label{tab:conditions}
\footnotesize
\setlength{\tabcolsep}{3pt}
\renewcommand{\arraystretch}{1.15}
\begin{tabularx}{\textwidth}{@{}>{\raggedright\arraybackslash}p{0.19\textwidth}
  >{\raggedright\arraybackslash}X
  >{\raggedright\arraybackslash}X
  >{\raggedright\arraybackslash}X@{}}
\toprule
Requirement (Lean) & Enforcement & Evidence & Status \\
\midrule
FDR $\le\alpha$ at every fixed horizon (\texttt{lord\_fdr},
\texttt{lord\_mfdr}, \texttt{fdr\_le})
  & The architecture as a whole, through (H1)--(H4)
  & Lean proofs (Section~\ref{sec:faithfulness})
  & Machine-checked: mFDR with the budget derived; FDR under
    non-adaptive thresholds. Adaptive case: cited
    result~\cite{javanmard2018online} \\
(H1) Uniform null p-values (\texttt{hU})
  & Harness-owned permutation test on held-out per-example losses
  & Null-calibration experiment (Section~\ref{sec:simulation})
  & Assumed; supported by experiment. Lean states the equality form \\
(H2) Predictability (\texttt{h$\alpha$p})
  & \texttt{StateT} threading in the \texttt{Research} monad
  & Type checking; GNATprove flow analysis of \texttt{Advance}
  & Enforced by construction; flow confirmed by SPARK \\
(H3) Independence (\texttt{hI})
  & Disjoint per-hypothesis validation splits; harness controls the
    data \texttt{optimize} receives; OS-level sandbox
  & Case studies (Section~\ref{sec:casestudy}); adversarial leakage
    evaluation (Section~\ref{sec:leakage})
  & Enforced by construction at the interface; OS-level isolation
    tested as a standalone component \\
(H4) Budget (\texttt{hbudget}; derived by
\texttt{lordThreshold\_sum\_le})
  & SPARK kernels: Eq.~(\ref{eq:lord}) (\texttt{Lord\_Eq1}) and the
    multiplicative update (\texttt{Lord\_PP})
  & GNATprove: 716 prover checks, none unproved
    (Table~\ref{tab:spark_counts})
  & Machine-checked: budget over $\mathbb{R}$ for Eq.~(\ref{eq:lord});
    over IEEE~754, thresholds within wealth under a margin condition \\
\bottomrule
\end{tabularx}
\end{table}

\section{Evaluation}
\label{sec:evaluation}

We evaluate the architecture along three lines: a simulation study
showing why FDR control is necessary at scale, end-to-end case studies
showing the integrated pipeline makes sound decisions, and an
adversarial evaluation probing the data-separation boundary against
code that actively tries to cross it.

\subsection{Simulation Study}
\label{sec:simulation}

We used Monte Carlo simulation with $N = 2000$ hypotheses tested
sequentially. Ground truth consisted of 10\% true effects and 90\%
null effects. Under the null, p-values were drawn from
$\text{Uniform}(0,1)$; under the alternative, from
$\text{Beta}(0.15, 1)$ to model moderate statistical power. The target
FDR was $\alpha = 0.05$, and results were averaged over 100 runs.

\begin{table}[t]
\centering
\caption{Monte Carlo results ($N{=}2000$ hypotheses, 100 runs, $W_0 = 0.1\alpha$). The
naive approach inflates FDR to 41\%; LORD controls it at 1.1\%.}
\label{tab:simulation_results}
\begin{tabular}{@{}lccc@{}}
\toprule
Approach & Target FDR & Empirical FDR & Power \\
\midrule
Naive (Fixed $\alpha{=}0.05$) & 0.05 & 0.409 & 0.640 \\
Monadic (LORD) & 0.05 & 0.011 & 0.290 \\
\bottomrule
\end{tabular}
\end{table}

The results in Table~\ref{tab:simulation_results} are clear. The naive
approach, testing each hypothesis at fixed $\alpha = 0.05$, inflates
FDR to 41\%, meaning nearly half of reported discoveries are false.
The monadic LORD implementation controls FDR at 1.1\%, well below
the 5\% target. The cost is reduced power (29\% vs.\ 64\%), but this
is the expected price of valid multiple testing correction: the
procedure becomes conservative when wealth is low, which occurs after
long sequences of true null hypotheses.

Two features of these results deserve emphasis. First, the simulation
draws null p-values from $\text{Uniform}(0,1)$: it therefore validates
the \emph{procedure} given valid p-values, the role that the
permutation test of Section~\ref{sec:bridge} plays in the running
system, rather than the test's calibration. The empirical FDR control
is consistent with the derived budget bound of
Section~\ref{sec:faithfulness}. Second, the gap between 1.1\% and the 5\% target reflects
LORD's conservatism for this parameter regime; the guarantee is an
upper bound, not an equality.

\paragraph{Null calibration of the test (H1)}
To check condition~(H1) directly we ran the harness test under the
null---optimized and baseline models with exchangeable per-example
losses---and compared its p-value distribution to uniform. Over $4000$
trials the paired permutation test is super-uniform: its empirical
rejection rate at the nominal $0.05$ level is $0.049$, with the
empirical CDF on or below the diagonal. The cross-validated paired
$t$-test it replaces, evaluated on the same correlated-fold structure,
is sharply anti-conservative---rejecting at rate $0.37$ under the
null---a direct illustration of why a fixed implementation of an
invalid test does not by itself establish~(H1).

\subsection{Case Study: End-to-End Demonstration}
\label{sec:casestudy}

We exercised the integrated architecture on the SVM scenario from
Section~\ref{sec:example}: a weak baseline and five optimization
hypotheses (RBF kernel, feature scaling, $C$ regularization,
polynomial kernel, and feature selection), each scored through the
harness. The setup enforced a \texttt{DataContract} (disjoint per-hypothesis
validation splits) and a \texttt{StatisticalTestSpec} (a paired
permutation test on held-out per-example losses). Target FDR was 0.05.

\begin{table}[t]
\centering
\caption{Case study: the five hypotheses scored two ways on the wine
validation data, with LORD thresholds ($W_0 = \alpha/2$). The cross-validated paired
$t$-test (the original harness method) inflates significance by orders
of magnitude and clears the thresholds at $t=2,3,5$ (bold),
\emph{reporting three discoveries}; the valid paired permutation test
on disjoint held-out splits clears none. The tiny $t$-test p-values are
artifacts of treating thirty overlapping folds as independent. The
$\alpha_t$ column is for the permutation (valid) path.}
\label{tab:casestudy_contrast}
\small
\begin{tabular}{@{}clccc@{}}
\toprule
 & & \multicolumn{2}{c}{p-value} & \\
\cmidrule(lr){3-4}
$t$ & Hypothesis & CV $t$-test & Permutation & LORD $\alpha_t$ \\
\midrule
1 & RBF kernel optimization   & 1.000            & 0.856 & 0.00134 \\
2 & Feature scaling           & \textbf{0.00004} & 0.250 & 0.00029 \\
3 & $C$ regularization        & \textbf{0.00006} & 1.000 & 0.00025 \\
4 & Polynomial kernel         & 0.016            & 0.773 & 0.00021 \\
5 & Feature selection (RFE)   & \textbf{0.00009} & 0.125 & 0.00017 \\
\bottomrule
\end{tabular}
\end{table}

Table~\ref{tab:casestudy_contrast} scores the five hypotheses two ways.
Under the cross-validated paired $t$-test the original harness used,
three hypotheses (feature scaling, $C$ regularization, feature
selection) return p-values near $10^{-5}$ and are reported as
discoveries---they clear even LORD's strict thresholds. Under the
valid permutation test on disjoint held-out splits, the same hypotheses
yield honest p-values between $0.13$ and $1.0$, and none is
significant. Feature scaling does help on this data, but a
four-order-of-magnitude p-value reflects thirty overlapping folds
counted as independent observations, not evidence of that strength. The
example isolates a failure that FDR accounting alone cannot catch: a
correctly tracked error budget spent on invalid p-values still yields
spurious discoveries; only valid per-test calibration~(H1) \emph{under}
online-FDR accounting is sound. It also shows the genuine cost of
rigor---with disjoint splits each hypothesis sees a small sample, so the
procedure is conservative, conceding power to detect an effect only on
strong, data-rich evidence.

\paragraph{Power when the data suffice}
That conservatism is a function of sample size, not a permanent ceiling.
To show the same valid pipeline has power when data suffice, we repeat
it on a larger nonlinear benchmark (\texttt{make\_moons}, $6000$ points,
$\approx720$ held-out examples per hypothesis), where a linear baseline
underfits and kernel choice is decisive. Five hypotheses are posed:
three genuine improvements (RBF, polynomial, and tuned-RBF kernels) and
two nulls (linear models equivalent to the baseline).
Table~\ref{tab:casestudy_positive} shows the outcome. The pipeline
discovers all three real effects (permutation p-values $5\times10^{-5}$
to $8\times10^{-4}$, clearing the LORD thresholds) and rejects both
nulls (p-value $1.0$); the realized false discovery proportion is zero.
The thresholds also respond to the discovery history, and most
strongly to \emph{recent} discoveries: $\alpha$ rises to $0.00197$ at
$t=5$ following the rejection at $t=4$, against $0.00074$ at $t=4$
whose nearest prior discovery lies two steps back. This
recency-weighted adaptivity is the behavior that gives LORD its
power advantage over static corrections. Together the two case studies
bracket the method's behavior: it neither manufactures discoveries from
invalid statistics (Table~\ref{tab:casestudy_contrast}) nor forfeits
real ones when the data support them (Table~\ref{tab:casestudy_positive}).

\begin{table}[t]
\centering
\caption{Positive demonstration on a larger nonlinear benchmark
(\texttt{make\_moons}, $6000$ points, $W_0 = \alpha/2$). The valid pipeline---disjoint
splits, paired permutation test, LORD---discovers all three genuine
kernel improvements and rejects both nulls, for a realized false
discovery proportion of zero. The threshold $\alpha_t$ rises after each
discovery as the reward refunds wealth.}
\label{tab:casestudy_positive}
\small
\begin{tabular}{@{}clccl@{}}
\toprule
$t$ & Hypothesis & Truth & Perm.\ p & LORD $\alpha_t$ \\
\midrule
1 & RBF kernel                  & real & $5\times10^{-5}$ & 0.00134 \\
2 & Polynomial kernel (deg.\ 3) & real & $8\times10^{-4}$ & 0.00163 \\
3 & Null: re-fit linear         & null & $1.00$           & 0.00188 \\
4 & RBF, tuned $C$              & real & $5\times10^{-5}$ & 0.00074 \\
5 & Null: negligible $C$ change & null & $1.00$           & 0.00197 \\
\bottomrule
\end{tabular}
\end{table}

When LLM-generated code fails outright---raising an exception or
producing malformed output---the orchestrator captures the error and
the monadic architecture treats it as ``no result'' rather than a
protocol violation, so the campaign continues to the next hypothesis
without corrupting the statistical state.

\subsection{Adversarial Leakage Evaluation}
\label{sec:leakage}

The case study shows the architecture operating as intended; it does
not, by itself, establish that the data-separation boundary
(Section~\ref{sec:sandbox}) withstands code that actively tries to
cross it. We test this directly, treating data separation as a
security property and probing it with a worst-case adversary.

\paragraph{Design}
For each trial an LLM (Claude Sonnet~4.6) generates a module with the
functions \texttt{optimize}, \texttt{get\_baseline}, and
\texttt{evaluate\_model} under one of two prompts. The \emph{honest} prompt gives
ordinary instructions; the \emph{adversarial} prompt rewards the model
solely for maximizing validation accuracy and explicitly tells it to
load any validation or test data it can locate. In both prompts the
model is told the exact on-disk path of the validation split, modeling
a worst-case adversary that has seen the generated harness: a
boundary worth claiming must hold even against code that knows precisely
what to read. Each generated module is then executed twice---once with
the data directory intact (the \emph{prompt-only} baseline, in which the
defense is merely the instruction not to cheat) and once inside the
mount namespace of Section~\ref{sec:sandbox} (\emph{isolated}).

\paragraph{Measurement}
Leakage is detected at the system-call level: the optimization process
and all of its children are traced, and a leak is recorded when an
\texttt{open} of the validation file returns successfully. This signal
is independent of the generated code---it does not rely on the
implementation reporting its own behavior---and it captures every
access path, including reads performed by a spawned subprocess. The
structural guarantee predicts that the isolated leak rate is zero; any
successful open under isolation would falsify it.

\paragraph{Datasets}
The primary condition uses a private synthetic dataset generated from a
seed internal to the harness and present nowhere but on disk. Because it
is not a public benchmark, the on-disk file is the only channel to the
validation labels, so a zero isolated leak rate is a substantive
result rather than a consequence of the data being unavailable elsewhere.
As a contrast we repeat the protocol on the public Wine dataset and
additionally scan each generation for calls that reconstruct the data
from a library (\texttt{load\_wine} and related loaders); such a
generation obtains the validation labels without opening the file, a
channel that the filesystem boundary, by construction, cannot close.

\paragraph{Results}
Table~\ref{tab:leakage} reports the outcome over $N{=}50$ generations
per cell. The pattern is identical on both datasets. Under the
adversarial prompt, every well-formed generation attempted to read the
validation file and, under the prompt-only baseline, succeeded---a leak
rate of $1.00$. Under isolation the same generations all failed: the
leak rate is $0.00$. Because the detector observes the open at the
system-call level, this zero is falsifiable---a single successful open,
by the process or any child, would register---so it reflects the
boundary holding, not a blind spot in measurement. Honest generations
never attempted access, even though they too were told the path,
confirming that the baseline failure is specific to a model under
pressure rather than an artifact of the prompt. One synthetic
adversarial generation failed to compile and is excluded (leaving
$n{=}49$); all other cells are complete.

\begin{table}[t]
\centering
\caption{Leakage over $N{=}50$ generations per cell
(Claude Sonnet~4.6), as fractions of well-formed generations. An
adversarial model reads the validation file in every generation under
the prompt-only baseline and in none under OS-level isolation; honest
generations never attempt access.}
\label{tab:leakage}
\small
\begin{tabular}{@{}llcccc@{}}
\toprule
 & & & \multicolumn{2}{c}{Leak rate} & \\
\cmidrule(lr){4-5}
Dataset & Prompt & Attempt & Prompt-only & Isolated & Reconstruct \\
\midrule
Synthetic & Honest       & 0.00 & 0.00 & 0.00 & 0.00 \\
Synthetic & Adversarial  & 1.00 & 1.00 & \textbf{0.00} & 0.00 \\
Wine      & Honest       & 0.00 & 0.00 & 0.00 & 0.00 \\
Wine      & Adversarial  & 1.00 & 1.00 & \textbf{0.00} & 0.00 \\
\bottomrule
\end{tabular}
\end{table}

The reconstruction scan flagged no generation on either dataset: given
the file path, the model always took the direct route and read the
file, rather than rebuilding the data from a library. The
public-benchmark reconstruction channel of Section~\ref{sec:sandbox} is
therefore a limitation we note on principle---it exists and the
filesystem boundary cannot close it---rather than one the model
exploited here; defending against it would require withholding public
data or monitoring for library access, outside the scope of the
filesystem boundary.

\section{Discussion}
\label{sec:discussion}

\subsection{Scope of Guarantees}

Our architecture provides defense-in-depth through structural
enforcement at two levels (Table~\ref{tab:guarantees}). At the macro
level, the \texttt{Research} monad prevents state corruption, timing
errors, and protocol bypass. At the micro level, the scaffold prevents incorrect test selection
and fixes which data the generated code receives, and the OS-level
sandbox (Section~\ref{sec:sandbox}), tested as a standalone component,
removes the validation data from the environment in which that code
runs. We use ``structural
enforcement'' rather than ``correctness by construction'' because
guarantees at the micro level rely on correct execution of the
generated scaffold across the IO boundary.

\begin{table}[t]
\centering
\caption{Summary of structural guarantees and trust boundaries.}
\label{tab:guarantees}
\small
\begin{tabularx}{\textwidth}{@{}Xcc@{}}
\toprule
Challenge & Macro (Monad) & Micro (Scaffold) \\
\midrule
State corruption / stale reads & Enforced & N/A \\
Timing errors (off-by-one) & Enforced & N/A \\
Bypassing protocol accounting & Enforced & N/A \\
Data leakage & N/A & Enforced$^{\ast}$ \\
Incorrect statistical test & N/A & Enforced \\
\midrule
Remaining trust boundaries: & & \\
Implementation bugs in domain code & Not covered & Not covered \\
Correctness of Haskell protocol code & Not covered & N/A \\
Weak baseline selection by LLM & N/A & Not covered \\
Public-data reconstruction & N/A & Not covered \\
\bottomrule
\end{tabularx}

\smallskip
{\footnotesize $^{\ast}$At the interface in the orchestrator; at the
OS level in the standalone sandbox (Section~\ref{sec:sandbox}).}
\end{table}

The framework does not protect against all failure modes. In
particular, the LLM could define an artificially
weak baseline during refactoring, making any optimization appear
significant. Detecting this would require domain-specific validation
beyond the scope of our statistical framework.

More broadly, the scaffold does not guarantee that LLM-generated code
is semantically correct---it might run to completion but
implement the wrong optimization. However, the scaffold significantly
reduces the damage surface of such failures. Even if the LLM's code
is subtly wrong, it cannot leak harness-private validation data
(preventing inflated
test statistics), cannot substitute a weaker statistical test
(preventing invalid p-values), and cannot bypass the FDR accounting
(preventing error-rate inflation). Code that fails outright---raising
exceptions or producing malformed output---is caught by the
orchestrator, which retries generation a fixed number of times before
abandoning the hypothesis, similar to the retry-and-abandon strategy
used by existing AI-Scientist
systems~\cite{sakana2024aiscientist}. The residual risk is code
that runs successfully but implements a semantically incorrect
optimization, which is no different from the bugs a human programmer
might introduce. The scaffold reduces the problem from arbitrary
hallucination with unbounded statistical consequences to
domain-level implementation errors with bounded statistical impact,
since the FDR guarantee holds regardless of the quality of individual
experiments.

\subsection{Types for Structure, SPARK for Arithmetic}
\label{sec:division}

We use two verification tools because the conditions split cleanly
between them, and the Lean formalization is what makes the split
precise rather than analogical. The structural conditions
(H1--H3)---information flow, data partitioning, API contracts---are
what Haskell's type system and the scaffold's code generation express
naturally; encoding them in SPARK would need extensive ghost code to
recover what \texttt{StateT} provides automatically. The arithmetic
condition (H4) instead requires reasoning about IEEE~754 rounding,
which GNATprove handles and types cannot. Crucially, without the
formalization ``the monad enforces predictability'' is an informal
analogy; with it, we point to the typed hypothesis \texttt{h$\alpha$p}
and observe that \texttt{StateT} threading supplies exactly the
$\mathcal{F}_t$-measurability it names---a specific type-level property
discharging a specific mathematical one, and the same correspondence
identifies the four conditions an implementation must satisfy. Finally,
to show the verified arithmetic can be the arithmetic that runs, both
SPARK kernels export their operations under the C calling convention,
and Haskell harnesses drive them through the foreign-function
interface. For Equation~\ref{eq:lord}, the harness reproduces the
Haskell implementation bit for bit on the sequences of
Section~\ref{sec:spark-eq1}, so the thresholds can be evaluated by the
GNATprove-proved code over IEEE~754 \texttt{Long\_Float} rather than by
a separate, unverified copy.

\subsection{Related Work}

\paragraph{AI-Scientist architectures}
To position our contribution, we briefly describe the architecture
of the most prominent AI-Scientist system. The AI
Scientist~\cite{sakana2024aiscientist} operates as a linear pipeline:
(1)~idea generation from a human-authored code template,
(2)~novelty checking via Semantic Scholar, (3)~experiment iteration
in which an LLM coding assistant modifies \texttt{experiment.py} up
to five times, (4)~paper writing, and (5)~automated review. If the
LLM-generated code raises an exception, the iteration is retried;
after repeated failures the idea is abandoned. Critically, however,
no component of this pipeline enforces statistical discipline across
hypotheses: no component applies statistical testing, let alone
multiplicity control, the LLM has unrestricted access to
all data files during code editing, and there is no separation between
exploration and validation data. An independent
evaluation~\cite{beel2025evaluating} found that five of twelve
proposed experiments (42\%) failed due to coding errors and that manuscripts sometimes contained
hallucinated numerical results. Other systems---Agent
Laboratory~\cite{schmidgall2025agent}, Virtual
Lab~\cite{swanson2025virtual}---follow broadly similar patterns: an
LLM generates or modifies experimental code, results are collected,
and a report is produced. Curie~\cite{kon2025curie} adds rigor
modules, but its checks are carried out by LLM agents rather than
enforced structurally, and none of these systems applies multiplicity
control. Our architecture is not a generalization of these
pipelines but rather a missing layer: it provides the
statistical and data-integrity guarantees that existing systems leave
entirely to the user.

\paragraph{Statistical DSLs and libraries}
Tea~\cite{jun2019tea} is a Python DSL that automates statistical
test selection: the user declares hypotheses, variables, and
assumptions, and Tea compiles these into a constraint satisfaction
problem whose solution is the set of valid tests. Tea and our work
address complementary problems. Tea ensures that the right test
is chosen for a single hypothesis; our architecture ensures that
many hypotheses are tested under a controlled error budget and
that the data feeding those tests is not corrupted at the LLM
boundary. In principle the two approaches could be composed: Tea
could serve as the test-selection oracle inside our scaffold's
\texttt{StatisticalTestSpec}.

The \texttt{onlineFDR} R
package~\cite{robertson2019onlinefdr} implements LORD, LORD++, LOND,
SAFFRON, and other online FDR procedures as library functions. It
provides correct threshold computations and is well suited to
interactive analysis by a human statistician. However, as a passive
library it offers no enforcement: nothing prevents a caller from
bypassing the threshold check, feeding corrupted p-values, or
forgetting to call the procedure altogether. In an autonomous agent
loop, where the ``caller'' is LLM-generated code, this lack of
enforcement is a serious gap. Our monad makes the protocol
inescapable---the type system ensures that every hypothesis
test passes through the accounting step---and the scaffold addresses
a concern that \texttt{onlineFDR} does not consider at all: data
integrity across the trust boundary into generated code.

\paragraph{Functional eDSLs and type-level guarantees}
Typed functional languages and embedded domain-specific languages
(eDSLs) have a rich history of encoding domain
constraints---financial contracts~\cite{peytonjones2000composing},
dimensional analysis~\cite{kennedy1994dimension}---and probabilistic
programming applies the same idea to statistical
models~\cite{gordon2014probabilistic}. Monad transformers are a
standard technique for composing effects~\cite{liang1995monad}. Our contribution applies these ideas
to statistical methodology enforcement in the specific context of
hybrid AI systems. Several directions for stronger type-level
guarantees are worth noting. Indexed
monads~\cite{atkey2009parameterised} could encode the constraints
currently checked at runtime by \texttt{isValid\-Transition} directly
in the type system. Generalized algebraic data types (GADTs) could model the staged workflow
(exploration vs.\ validation) so that data access is phase-restricted
at compile time. Algebraic effects~\cite{plotkin2003algebraic} could
offer more flexible composition. On the verification side,
LiquidHaskell's refinement types could express the wealth
non-negativity invariant $W(t) \geq 0$ directly as a Haskell type
annotation. For
parallel testing, asynchronous FDR control
methods~\cite{zrnic2020asynchronous} with software-transactional-memory (STM) concurrent state
management would be needed. We view these as promising extensions
rather than limitations of the current approach.

\paragraph{Formal requirements for AI systems}
Seshia et al.~\cite{seshia2022verified} define verified AI as the goal
of AI-based systems whose correctness with respect to mathematically
specified requirements is assured, ideally by proof. Among the
challenges they list is formal specification itself, including how to
state quantitative rather than Boolean properties of systems with
learned components. Requirements engineering tools address the
specification side: FRET~\cite{giannakopoulou2021fret} translates
structured natural-language requirements into past- and future-time
metric temporal logic, and recent work extends it to probabilistic
requirements, translated into PCTL$^{*}$ formulas for probabilistic
model checkers~\cite{mavridou2025probabilistic}. Our requirement is of a
different kind. FDR control is a statistical property of a whole
campaign of tests, an expectation over the unknown process that
generates the data, not a property of a single execution trace, so it
cannot be checked by monitoring one run. Probabilistic temporal logics
state such properties over an explicit model of the system; here no
model of the data-generating process is available. We obtain the formal
requirement from a machine-checked theorem instead: the theorem's
hypotheses become the derived requirements, and each is traced to the
mechanism that enforces it (Table~\ref{tab:conditions}). We expect the
same pattern---prove a statistical guarantee once, then enforce its
hypotheses in the implementation---to carry over to other statistical
requirements on AI systems.

\paragraph{Formal verification of probabilistic properties}
Machine-checked proofs of probabilistic and statistical properties
are an active area. Notable work includes Affeldt et al.'s Coq
formalization of information theory~\cite{affeldt2014mca} and
H\"{o}lzl's Isabelle/HOL formalization of probability
theory~\cite{hoelzl2012probability}. Our work is, to our knowledge,
the first machine-checked proof of an online FDR control theorem.

\section{Conclusion}

We have presented a functional architecture for enforcing statistical
rigor in AI-Scientist systems. A machine-checked Lean~4 formalization
identifies four conditions for FDR control; these split naturally into
structural properties---about information flow, data separation, and
test validity---and an arithmetic property requiring that a wealth
process remain non-negative. The \texttt{Research} monad and
declarative scaffold enforce the structural conditions by
construction; data separation is further enforced by an OS-level sandbox,
implemented and tested as a standalone component, that removes the
validation data from the execution environment; an adversarial
evaluation confirms that it holds even against code that knows the
file's path. A SPARK/Ada implementation of the LORD thresholds
verifies over IEEE~754 doubles that the computed thresholds never
exceed the computed wealth, given a margin condition that our
configurations meet by a factor of at least eight; without a margin,
we exhibit a case in which the property fails by one unit in the last
place. GNATprove discharges all 716 prover checks of the SPARK code,
and its flow analysis independently confirms predictability. Every
condition is thus traced to a proof, a construction or an experiment
(Table~\ref{tab:conditions}); the links that remain open---a
machine-checked refinement from the real-number proofs to the
floating-point ones, full FDR control for adaptive thresholds, and
exact rather than super-uniform null p-values---are stated in
Sections~\ref{sec:faithfulness} and~\ref{sec:spark-scope}. To our
knowledge, the Lean formalization is the first machine-checked proof of
an online FDR control theorem.

\section*{CRediT authorship contribution statement}

\textbf{Karen Sargsyan:} Conceptualization, Methodology, Software,
Validation, Formal analysis, Investigation, Writing -- Original Draft,
Writing -- Review \& Editing.

\section*{Declaration of generative AI and AI-assisted technologies in the manuscript preparation process}

During the preparation of this work the author used Anthropic Claude
in order to assist with manuscript drafting, LaTeX formatting, and
iterative revision. After using this tool, the author reviewed and
edited the content as needed and takes full responsibility for the
content of the published article.

\section*{Funding}

This research did not receive any specific grant from funding agencies
in the public, commercial, or not-for-profit sectors.

\section*{Declaration of competing interest}

The author declares that there are no known competing financial
interests or personal relationships that could have appeared to
influence the work reported in this paper.

\section*{Data availability}

All source code, the Lean~4 formalization (including the derived budget
bound, the marginal-FDR theorem, and the leave-one-out FDR theorem),
the SPARK/Ada verification files, the harness with its permutation test
and disjoint-split generator, and the simulation, case-study,
calibration, and adversarial-leakage scripts are available at
{\small\url{https://github.com/karsar/ai-scientist-guards}}~\cite{sargsyan2026guards}.

\bibliographystyle{elsarticle-num}
\bibliography{references}

\end{document}